\documentclass{aa}
\usepackage{graphicx}
\usepackage{txfonts}
\usepackage{amsmath}
\usepackage{graphics}
\usepackage[dvips]{color}

\begin{document}

   \title {The 2175 \AA~Bump Features  in FeLoBAL Quasars:  \\ One Indicator of MW-like Dust in the  Nuclear  Region of Quasars}

    \titlerunning {MW-like Dust in FeLoBALQs}
    \authorrunning {Zhang et al.}

   \author{Shaohua Zhang \inst{1},
            Jian Ge \inst{2},
            Tuo Ji \inst{3},
            Peng Jiang \inst{3},
            Zhijian Luo \inst{1},
            Xiang Pan \inst{3},
            Xiheng Shi \inst{3},
            Chenggang Shu \inst{1},
            Jianguo Wang \inst{4},
            Hubing Xiao \inst{1},
          \and
            Hongyan Zhou\inst{3,5,6}
          }

   \institute{  Shanghai Key Lab for Astrophysics, Shanghai Normal University, Shanghai 200234, China
         \and    Key Laboratory for Research in Galaxies and Cosmology, Shanghai Astronomical Observatory, Chinese Academy of Sciences, Shanghai 200030, China
         \and    Key Laboratory of Polar Science, Polar Research Institute of China, Ministry of Natural Resources, Shanghai 200136, China
         \and    Key Laboratory for the Structure and Evolution of Celestial Objects, Yunnan Observatories, Chinese Academy of Sciences, Kunming, Yunnan 650011,  China
         \and    Key Laboratory for Research in Galaxies and Cosmology, Department of Astronomy, University of Science and Technology of China, Chinese Academy of Sciences, Hefei, Anhui 230026, China
         \and   School of Astronomy and Space Science, University of Science and Technology of China, Hefei, Anhui 230026, China
             }

   \date{Received xx xx, xxxx; accepted xx xx, xxxx}

  \abstract
   {To investigate the properties of dust in the nuclear region of quasars, we explored the extinction curves of the iron low-ionization broad absorption line (FeLoBAL) quasar SDSS J163004.29+311957.6 and its two  analogues. The parameterized extinction curves indicated the Milky Way-like 2175 \AA~ bump features in underlying extinction, which are similar to those seen in the Local Group and a subset of high-redshift star-forming galaxies. Compared to the bump features in the Large Magellanic Clouds (LMC), the detections in this work are much closer to those in the Milky Way (MW). These bump features, as well as those in the high- and low-ionization broad absorption line (BAL) quasars of Zhang et al., are probably the counterpart of the 2175 \AA~  bump features in the  quasar environment. This type  of dust grain is generally small, easily disrupted by high-energy photons and   has difficulty surviving in the radiation field of the active galactic nucleus (AGN). However, due to the presence of absorption-line  outflows, the 2175 \AA~  bump feature in quasars, which should be rare, is seen many times in BAL quasars.  The shielding effect of outflow clouds allows the MW-like dust grains to be assembled or   extends  the survival period in the quasar nuclear region. The process, and physical and chemical conditions  deserve further observational study and investigation. }


  \maketitle

\section{Introduction}

Dust is everywhere in the universe,  and although this dust is tiny, it plays an important role in many astrophysics processes.
Absorption, scattering, and  re-radiation of the dust reshape the spectral energy distribution (SED) of galaxies and active galactic nuclei (AGNs).
At the same time, as an important component of the interstellar medium (ISM),   the  dust grain surface offers a place to efficiently form molecules as a critical step for  star formation,  which creates galaxies (Draine 2003).
Understanding dust properties and their evolution is crucial to providing complete knowledge of ISM processes,  star formation  and galaxy evolution.

In the galaxy, dust fills interstellar space. Its distribution is heterogeneous, and it is known to be clumpy with cold, dense regions of star formation.
In the AGN, the dust generally accumulates on the periphery of the accretion disk and forms a circumnuclear dusty ``torus'' as an important structural element in the AGN ``unified'' schemes (e.g., Antonucci 1993; Urry \& Padovani 1995).
The torus obscures and absorbs  high-energy  photons emitted from the central engine and re-radiates at thermal infrared (IR) wavelengths (e.g., Rees et al. 1969; Barvainis 1987)  ; thus, it is also considered to be the reason  for  the appearance of Type I/II AGNs.
Beyond that, there is still a large amount of dust distributed in various parts of the nuclear region, such as the AGN outflows.
  Studies have found that  high-speed outflows have an  unexpectedly  significant correlation with hot dust emission, which suggests  a dust outflow scenario (e.g., Wang et al. 2013; Zhang et al. 2014).

Dust attenuation, polarimetry, and thermal radiation are all efficient observational means for  dust.
The blocking effect on centre  photons implies the existence of a dusty torus, and polarimetric and near-/mid-IR spectroscopic/photometric observations have presented  abundant information on the physical and structural characteristics (e.g., Krolik \& Begelman 1988; Pier \& Krolik 1992; Dullemond \& van Bemmel 2005; Tristram et al. 2007; Nenkova et al. 2008; Alonso-Herrero et al. 2011; Tristram et al. 2014; Garc{\'{\i}}a-Gonz{\'{a}}lez et al. 2017; Zhuang et al. 2018; Lopez-Rodriguez et al. 2020; Lyu \& Rieke 2021).
Relatively speaking, dust attenuation can delve into the details of interstellar dust grains, such as the size distribution of dust grains (e.g., Cardelli et al. 1989; Whittet 2003) and even the chemical compositions (e.g., Li \& Draine 2001; Draine 2003).
Furthermore, measuring the extinction curve and making a corresponding reddening correction allow  accurately deriving of the intrinsic SED, luminosity functions, and physical properties of AGNs.

  Regarding  the AGN extinction law, Czerny et al. (2004) compared the composites of red and blue quasars, and Gaskell et al. (2004) also  checked  the composites of quasars with different radio properties. 
  The above composite comparative studies found that quasar extinction curves are, on average, significantly flatter in ultraviolet (UV)   than those of local galaxies, and they do not show any traces of the 2175 \AA~ bump feature, which is  ubiquitous in  extinction curves of the Milky Way (MW) and Large Magellanic Clouds (LMC) (see Draine 1989).
  Similar results have been  obtained from the analyses of the quasar colour distribution (Richards et al. 2003; Hopkins et al. 2004), and the red ``tail'' of the colour distribution of SDSS quasars is well described by the Small Magmatic Cloud (SMC) like reddening.
However, Jiang et al. (2013) reported the quasar intrinsic anomalously steep reddening law via the exceptional source IRAS 14026+4341,
and the extinction curves detected in all individuals of the high $A_{\rm V}$ quasar sample are  also steeper than the SMC law (Zafar et al. 2015).
In the study of polar dust obscuration in nearby broad-line active galaxies (BLAGNs) from the XMM-XXL field, Buat et al. (2021) reported that even 60\% of sources containing polar dust exhibit a steep extinction curve.  
Indeed, the reality is probably more complex with a mix of different shapes (e.g., Buat et al. 2021), and we are still far from understanding quasar extinction phenomena.
Moreover, Zhang et al. (2017) derived the SMC-type reddening law of a dusty torus in three narrow-line Seyfert 1 galaxies (NLS1s).
  However,  the extinction-to-gas ratios ($E{\rm (B-V)}/N_{\rm H}$) of the two cases are comparable to the Galactic standard value (Bohlin et al. 1978) and approximately ten to twenty times larger than that of the SMC (Weingartner \& Draine 2001).
These findings indicate the dominance of small dust grains in the quasar environment , which offers the possibility of detecting the 2175 \AA~ bump feature in quasars.

  To date, dozens of 2175 \AA~ bump features have been reported at high redshifts.   These detections are mostly found in individual intervening  systems  towards background quasars, including metal (e.g., Mg II, Zn II, C I, etc.) absorbers (e.g., Wang et al. 2004; Srianand et al. 2008; Noterdaeme et al. 2009; Zhou et al. 2010; Jiang et al. 2010a,b, 2011; Ledoux et al. 2015; Ma et al. 2017) and 
DLAs  (e.g., Wucknitz et al. 2003; Junkkarinen et al. 2004; Prochaska et al. 2009; Wang et al. 2012; Ma et al. 2015) or towards gamma ray burst (GRB) afterglows (e.g., Ellison et al. 2006; Vreeswijk et al. 2006; Prochaska et al. 2009; Liang \& Li 2010; Schady et al. 2012; Zafar et al. 2011, 2012, 2018). These systems have a high metallicity and heavy dust depletion level (e.g., Ma et al. 2017).
These detections reveal information about the ISM in the foreground galaxies of quasars/GRBs or GRB hosts rather than the quasar itself. The quasar-associated 2175 \AA~  bump features  are still rarely detected,  and only  in a limited number of quasar intrinsic extinction curves (Zhang et al. 2015a; Pan et al. 2017; Shi et al. 2020).
Theoretically, the carriers of 2175 \AA~  bump features are  small-sized carbonaceous and silicate grains (e.g., Joblin et al. 1992; Weingartner \& Draine 2001; Li \& Draine 2001; Steglich et al. 2010; Blasberger et al. 2017)  ; however, they are easily disrupted by high-energy photons have difficulty existing in the radiation field of the centre engine (e.g., Voit 1992 and reference therein).
Why can this type of dust grain be produced and survive in this harsh environment? This is an interesting topic worth further investigation.
Exploring individuals, sample expansion and further detailed studies of important targets may provide a possibility for researching the formation and destruction of 2175 \AA~ dust grains in the nuclear region of quasars.

Broad absorption lines (BALs) are the most famous quasar intrinsic absorption systems, and they are one of the direct  performances  of AGN outflows. They appear in the spectra of $\sim 10 - 26\%$ of optically selected quasars and are often observed as absorption by the ions of C IV, Si IV, Al III and Mg II (e.g., Weymann et al. 1991; Reichard et al. 2003; Trump et al. 2006; Gibson et al. 2009; Scaringi et al. 2009; Zhang et al. 2010; Allen et al. 2011).
Moreover, iron low-ionization broad absorption lines (FeLoBALs) are a rare and special subclass of BALs, which exhibit absorption from excited fine-structure levels or excited atomic terms of Fe II or Fe III (e.g., Hazard et al. 1987; Becker et al. 1997, 2000; Menou et al. 2001; Hall et al. 2002; Zhang et al. 2015b). The mid-IR spectroscopic observation of 6 FeLoBAL quasars detected unidentified emission signatures of carbonaceous molecules (Farrah et al. 2010), which implies that the carriers of 2175 \AA~  bump features can probably survive in the environments of BAL quasars. 
In Zhang et al. (2015a), the initial exploration of 2175 \AA~  bump features  among Mg II absorption line systems on quasar spectra in the SDSS DR10 led to the identification of excess broadband extinction near 2250 \AA~ in 7 BAL quasars. These cases are in the redshift range of   $1.78$ to $2.29$, and 5 of them show low-ionization BALs (LoBALs) of Mg II and Al III.
This provides direct evidence of the coexistence of BALs and 2175 \AA~ bump features in quasars. 

  In this work, we report the detection of the 2175 \AA~ bump feature in 3 FeLoBAL quasars.
The remainder of this paper is organised as follows: in Sect. 2, we describe the observational data available and data reduction. In Sect. 3, we present the derivation of the extinction curve, Sect. 4 provides a discussion, and Sect. 5 summarises the main conclusions.

\section{Observations and data reduction}

After Zhang et al. (2015a), 2175 \AA~ bump features in the spectra of the SDSS DR7 FeLoBAL quasars (Shen et al. 2011) were systematically explored following the search procedure of Zhang et al. (2015a). 
The quasar composite reddened by a parameterised extinction curve (introduced in the first two paragraphs of Sect. 3) was used to fit the observed spectra in the quasar rest frame and obtain potential bump candidates, and then a simulation technique of the control sample (Jiang et al. 2010a,b; also see the fifth paragraph of Sect. 3) was performed to gauge the significance of the bump features. 10 bump candidates were screened out with 2175 \AA~ bump features at a $> 3 \sigma$ level of statistical significance (see Table 1).
After performing a visual examination, 7 ``false'' signals are rejected. This is simply because their spectral wavelength ranges only cover part of the bump feature, and no effective photometric data limit the far-UV models. We finally identified 3 credible candidates, i.e., J102036.10+602339.0 (referred to as SDSS J1020+6023 hereafter), SDSS J134951.93+382334.1 (referred to as SDSS J1349+3823 hereafter), and SDSS J163004.29+311957.6 (referred to as SDSS J1630+3119 hereafter).
Their emission-line redshifts given by the SDSS DR7 quasar catalogue (Schneider et al. 2010) are $0.9940\pm0.0006$, $1.0943\pm0.0009$, and $1.9960\pm0.0007$, respectively. 
Since SDSS J1630+3119 has the most observational data and its spectrum shows the entire bump profile, it is analysed and reported in the text in detail. 
For SDSS J1020+6023 and SDSS J1349+3823, their optical spectra cannot completely cover the bump profiles because of their lower redshifts; fortunately, the reliability of spectral fitting is verified by the GALEX data; thus, they are listed in the appendices. 

   \begin{table*}[tph]
                \tiny
                \caption{  Best Fitted Parameters of  10 bump candidates in  FeLoBAL quasars}\label{tab1}
                \centering
        $$
        \begin{array}{cc ccc cc c}
            \hline
            \hline
            \noalign{\smallskip}
             {\rm Name~ (SDSS J)} & z & c_1 &   c_2 &   c_3 &   x_0 &   \gamma  & {\rm Significance} \\
            \noalign{\smallskip}
            \hline
            \noalign{\smallskip}
101927.37+022521.4 & 1.3643 & -1.95\pm0.02 & 0.65\pm0.01 & 0.76\pm0.06 & 4.46\pm0.01 & 0.99\pm0.04 & 5.6\sigma\\
102036.10+602339.0 & 0.9940 & -1.82\pm0.02 & 0.54\pm0.01 & 0.95\pm0.08 & 4.49\pm0.01 & 1.20\pm0.03 & 6.9\sigma\\
102249.27+130125.0 & 1.2168 & -0.87\pm0.03 & 0.55\pm0.01 & 1.41\pm0.07 & 4.78\pm0.02 & 1.68\pm0.04 & 3.7\sigma\\
110711.40+082331.2 & 1.3875 & -1.73\pm0.04 & 0.81\pm0.01 & 0.57\pm0.04 & 4.69\pm0.01 & 0.95\pm0.05 & 6.2\sigma\\
112220.76+153927.8 & 1.0996 & -1.38\pm0.01 & 0.43\pm0.01 & 1.11\pm0.06 & 4.56\pm0.01 & 1.10\pm0.02 & 7.0\sigma\\
134951.93+382334.1 & 1.0943 & -0.73\pm0.02 & 0.37\pm0.01 & 2.02\pm0.13 & 4.67\pm0.01 & 1.18\pm0.08 & 17.2\sigma\\
142010.28+604722.3 & 1.3450 & -0.58\pm0.03 & 0.51\pm0.01 & 0.39\pm0.06 & 4.58\pm0.01 & 0.87\pm0.04 & 8.7\sigma\\
152438.79+415543.0 & 1.2299 & -1.34\pm0.01 & 0.41\pm0.01 & 0.70\pm0.06 & 4.47\pm0.01 & 0.96\pm0.03 & 10.9\sigma\\
155633.78+351757.3 & 1.4950 & -3.46\pm0.02 & 1.00\pm0.01 & 1.35\pm0.10 & 4.56\pm0.01 & 1.45\pm0.02 & 7.6\sigma\\
163004.29+311957.6 & 1.9960 & -3.37\pm0.05 & 1.11\pm0.01 & 0.90\pm0.07 & 4.51\pm0.01 & 0.92\pm0.07 & 31.2\sigma\\
           \noalign{\smallskip}
           \hline
           \hline
        \end{array}
        $$
   \end{table*}

SDSS J1630+3119 is an infrared-luminous quasar with an extremely faint UV emission. The broadband SED for SDSS J1630+3119 from UV to near-IR is presented by the photometric images taken with the SDSS (York et al. 2000) at  the $u'$, $g'$, $r'$, $i'$, and $z'$ bands, the Two Micron All Sky Survey (2MASS; Skrutskie et al. 2006) at  the  $J$, $H$, and $Ks$ bands, and the Wide-field Infrared Survey Explorer (WISE; Wright et al. 2010) at  the  $W1$, $W2$, $W3$, and $W4$ bands. The optical and near-IR photometric data of SDSS J1630+3119 join smoothly, suggesting that the quasar did not vary significantly between the SDSS and 2MASS observations. The multiband magnitudes are  summarised in Table 2.

   \begin{table}
		\tiny
		\caption{  Broadband photometric data of SDSS J1630+3119}\label{tab2}
		\centering
        $$
        \begin{array}{cccc}
            \hline
            \hline
            \noalign{\smallskip}
	     {\rm Band} & {\rm Magnitude} & {\rm Date-Observation} &	{\rm Survey}\\
            \noalign{\smallskip}
            \hline
            \noalign{\smallskip}
		u'   	&	23.15\pm0.46   	&	04/29/2003   &	{\rm SDSS} \\
		g'   	&	21.41\pm0.04   	&    04/29/2003   &	{\rm SDSS} \\
		r'    	&  	20.12\pm0.02   	&    04/29/2003   &	{\rm SDSS} \\
		i'    	&  	19.16\pm0.02  	&	04/29/2003   &	{\rm SDSS} \\
		z'   	&  	18.31\pm0.03   	&	04/29/2003   &	{\rm SDSS} \\
		J    	&  	17.420\pm0.221 &	04/05/1998   &	{\rm 2MASS} \\
		H    	&  	16.178\pm0.170 &	04/05/1998   &	{\rm 2MASS} \\
		Ks       &  	15.871\pm0.205 &	04/05/1998   &	{\rm 2MASS} \\
		W1   &  	15.217\pm0.205 &	02/17/2010   &	{\rm WISE} \\
		W2   &  	14.080\pm0.044 &	02/17/2010   &	{\rm WISE} \\
		W3   &  	10.720\pm0.076 &	02/17/2010   &	{\rm WISE} \\
		W4   &     8.191\pm0.206  &	02/17/2010   &	{\rm WISE} \\
		V   &	-   &	07/06/2005-10/22/2013   &	 {\rm CSS} \\
		g   &	-   &	06/17/2009-06/27/2013   &	 {\rm Pan-STARRS} \\
		r    &	-   &	06/07/2010-06/20/2013   &	 {\rm Pan-STARRS} \\
		i    &	-   &	06/09/2009-05/11/2014   &	 {\rm Pan-STARRS} \\
		z   &	-   &	03/01/2010-08/16/2013   &	 {\rm Pan-STARRS} \\
		y   &	-   &	02/20/2010-08/22/2013   &	 {\rm Pan-STARRS} \\
		zr  &	-   &	06/11/2018-09/02/2019   &	 {\rm ZTF} \\
           \noalign{\smallskip}
           \hline
           \hline
        \end{array}
        $$
   \end{table}
   \begin{figure}
   \centering
   \includegraphics[width=\hsize]{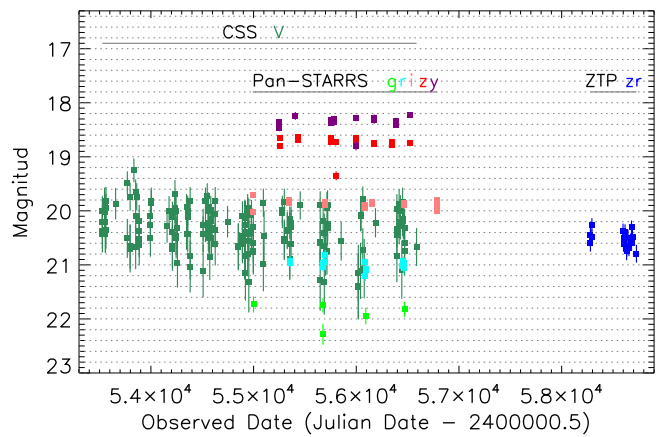}
   \caption{Light curves of SDSS J1630+3119 from  CSS, Pan-STARRS, and ZTF surveys. The raw data are shown with errors in different colours. The CSS continually monitored SDSS J1630+3119 for more than 8 years (170 epochs) at the $V$ band beginning on 6 July 2005,   in which the light curve means that there is no apparent change in the optical continuum within large measurement errors. Pan-STARRS  has performed another extensive monitor since 9 June 2009, and the magnitudes of 68 epochs confirm the latter half of the light curve observed by the CSS. After  approximately 5 years, SDSS J1630+3119 was monitored again by the ZTF and has 18 new observations with a $zr$ filter. The light curves with  a  high photometric accuracy from the Pan-STARRS and ZTF suggest that SDSS J1630+3119  does not have significant long-term variability in optical and near-infrared bands.} \label{fig1}
    \end{figure}

SDSS J1630+3119 was targeted for spectroscopic observations by the SDSS as   a high-redshift quasar candidate (\mbox{$LEGACY\_TARGET1 = QSO\_HIZ$}).
Two archival observations for SDSS J1630+3119 from the SDSS and SDSS-III/BOSS were performed on 4 April 2005 (DR7; Abazajian et al. 2009) and 7 July 2011 (DR16; Dawson et al. 2016), and the 1D spectra  were accessed from the SDSS Science Archive Server. 
The BOSS spectrum multiplied by  approximately  1.2 times is in excellent agreement with the SDSS spectrum.
Meanwhile, the Catalina Sky Surveys (CSS; Drake et al. 2009), the Panoramic Survey Telescope \& Rapid Response System (Pan-STARRS; Chambers et al. 2016), and the Zwicky Transient Facility (ZTF; Masci et al. 2018) present that SDSS J1630+3119 does not have significant long-term variability in  either  optical or near-IR bands within the current measurement errors for 14 years beginning on 6 July 2005 (Fig. 1).
Thus, the spectral flux density differences between the two spectroscopic observations probably come from the spectrophotometric calibration of the different SDSS Optical Spectroscopy Pipelines (Version `$26$' and `$v5\_13\_0$' ).
Moreover, the scaled BOSS spectrum agrees very well with the SDSS photometric data , which indicates that the recalibration is reliable and that SDSS J1630+3119 did not vary significantly between the SDSS photometric and spectroscopic observations.
In the following analysis, the scaled BOSS spectrum is used to derive the extinction curve because of its wider wavelength coverage.
All of the data have been corrected for the Galactic reddening of $E{\rm (B-V)}=0.032$ (Schlegel et al. 1998) and the Fitzpatrick (1999) reddening curve before  any  further analysis.

The spectrum of SDSS J1630+3119 shows unusual absorption features and has been  classified as a BAL quasar in Gibson et al. (2009) and Shen et al. (2011).
In Fig. 2, we present the observed spectrum of SDSS J1630+3119 and the absorption troughs of C IV, Al III, Mg II and iron multiplets (marked by grey dot-dashed lines and characters).
A typical FeLoBAL quasar with  absorption troughs similar to those of  SDSS J1630+3119 and SDSS J144002.24+371058.5 (referred to as SDSS J1440+3710 hereafter) is over-plotted for a comparison,  in which the spectrum is multiplied by a factor  to match  that of SDSS J1630+3119 in the wavelength ranges of $1600 - 1900$ \AA~ and $2800 - 3300$ \AA.
The high-ionization absorption lines of C IV $\lambda\lambda1548,1551$ in the spectrum of SDSS J1630+3119 are revealed by troughs near the centre of the corresponding lines, which have an extremely broad velocity width from approximately 4000 to $-13,000$ \mbox{km s$^{-1}$} with respect to the systemic velocity.
The numerous absorption lines from low-ionization species, such as Mg II, Al III, and Fe II, that  characterise  the spectrum show some different   appearances that  are narrower and heavily blended together for most of them in the lower-resolution spectrum.
In the figure, we identify some strong absorption troughs of Fe II multiplets UV 1, UV 2, UV 3, and UV 62,63 near 2600 \AA, 2382 \AA, 2344 \AA, and 2750 \AA~(in the quasar rest frame), and other weak multiplets of Fe II UV 8 and UV 44,45 on the longward of the C IV emission line, and Fe II UV 64, UV 145, and UV 263 on either side of the UV 1 multiplets are not marked.

   \begin{figure}
   \centering
   \includegraphics[width=0.95\hsize]{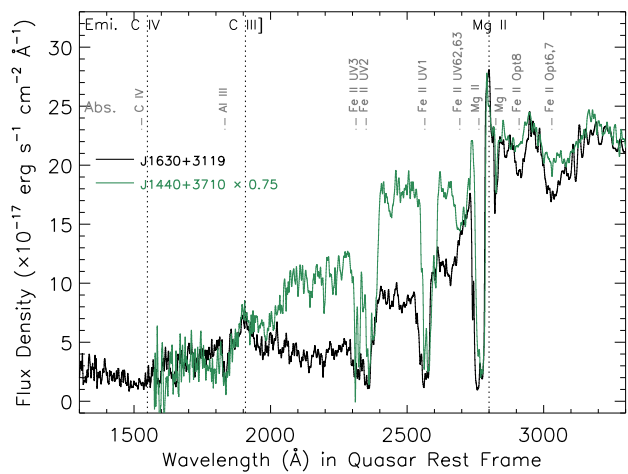}
   \caption{  Observed spectrum of SDSS J1630+3119 and  BAL troughs.
The spectrum was slightly smoothed with a Gaussian kernel with a sigma of 3 spectral pixels to help show the absorption features.
The spectrum of SDSS J1630+3119 is dominated by a strong broad and blueshifted absorption troughs from Mg II, Al III, C IV, and Fe II UV 1, UV 2, UV 3, UV 62,63, Opt. 6,7 and Opt. 8 multiplet transitions, the vertical dashed-dotted grey marks indicate the expected rest frame positions of the resonant lines in these features.
The comparison spectrum (shown in green) with similar absorption troughs to SDSS J1630+3119 from the FeLoBAL quasar SDSS J1440+3710 was smoothed and multiplied by 0.75   to match  that of SDSS J1630+3119 in the wavelength ranges of $1600 - 1900$ \AA~ and $2800 - 3300$ \AA. } \label{fig2}
    \end{figure}

\section{Extinction Curve and  Bump}

In Fig. 3(a), we present the observed data of SDSS J1630+3119, including the original spectrum (black curve) and multiband magnitudes (orange squares).
The dramatic flux drop shortward of the Mg II broad emission line (BEL) appears to slow down somewhere between the SDSS $r'$ and $i'$ bands, suggesting that a broad absorption bump occurs there.
In the MW and other galaxies in the Local Group, the interstellar extinction is most commonly obtained through the ``pair method'' by comparing the spectra of two stars of the same spectral type, one of which is reddened ($f_{\rm obs}$) while the other is unreddened ($f_{\rm model}$).
A similar technique, called the ``quasar spectrum pair method'', was used to derive the quasar extinction curve and identify 2175 \AA~ bump  features in the quasar spectrum (e.g., Wang et al. 2004; Zhou et al. 2010).  The reddened spectrum is  an observed quasar spectrum, and the unreddened  spectrum is replaced by the quasar composite or an individual blue quasar spectrum.
In this work, the unreddened spectrum is set to a quasar composite, which is obtained by combining the SDSS DR7 quasar composite ($\lambda < 3000$ \AA; Jiang et al. 2011) and the near-IR quasar template ($\lambda \ge 3000$ \AA; Glikman et al. 2006).

 We borrowed the Fitzpatrick \& Massa (1990) parametrization of the optical/UV extinction curve; 
  however,  the extinction curve we measured in this work is $A^*(x)$ rather than $k(\lambda-V)$ ($\equiv E{\rm (\lambda-V)}/E{\rm (B-V)}$, used in the original Fitzpatrick \& Massa formula).  
 The reason for this will be explained in the next paragraph. In fact, the above two parameters are equivalent for a given extinction curve.
 $A^*(x)$ is described as a combination of a linear background and a Drude component,  representing  the underlying extinction and the possible 2175 \AA~ bump feature, respectively,
\begin{eqnarray}
A^*(x) = {\rm -2.5 log} \frac{f_{\rm obs}}{f_{\rm model}}
= c_{1} + c_{2}x + c_{3}\frac{x^2}{(x^2-x_{0}^2)^2+x^2\gamma^2},
\end{eqnarray}
where $x~ (\equiv\lambda^{-1})$ is  the  inverse wavelength and $x_{0}$ and $\gamma$ are the peak position and full width at half maximum (FWHM) of the Drude profile, respectively. 
 The UV linear component is set by the slope $c_{2}$ and intercept $c_{1}$, which depend on the relative intensity of both model and observed spectra. In this work, the model  and observed spectra are not normalized. 
If they are normalized at the IR wavebands where the extinction  is negligible, the parameter $c_{1}$ will be zero. 
Meanwhile, the far-UV curvature term ($x\geq 5.9~ \rm \mu m^{-1}$) in  the Fitzpatrick \& Massa formula  is excluded. The limited wavelength coverages and the low signal-to-noise ratio (S/N) blueward spectra of SDSS J1630+3119  are insufficient for properly identifying the extinction in the far-UV band.
Here, the 2175 \AA~  bump feature  is supposed to have the same redshift as the quasar. If the devised peak position is much larger than the mean value of $x_{0}= 4.59\pm0.027$ $\mu$m$^{-1}$ in galactic interstellar curves, then the bump in the quasar spectrum would be the intervening system, which is not associated  with  the quasar.
After masking the wavelength range  with Mg II, Al III, C IV, and strong iron absorption troughs, the spectrum is fitted with models as the quasar composite that is reddened by dust with the parameterised extinction curve.
We perform the least squares minimization using  the Interactive Data Language (IDL) procedure MPFIT\footnote{The Markwardt IDL Library is available at http://cow.physics.wisc.edu/$\sim$craigm/idl/idl.html.} developed by Markwardt (2009). 
The best-fitting model yields $c_{1}=-3.37\pm0.05$, $c_{2}=1.11\pm0.01$, $c_{3}=0.90\pm0.07$, $x_{0}=4.51\pm0.01$ $\mu$m$^{-1}$, and $\gamma=0.92\pm0.07$ $\mu$m$^{-1}$. The uncertainties are the formal MPFIT errors.

In general, the  extinction  curves are exhibited in their native normalization ($E{\rm (\lambda-V)}/E{\rm (B-V)}$) or are normalised to an absolute scale ($A_{\lambda}/A_{\rm V}$). 
The visual/near-UV curves would be approximated by an analytic formula with one parameter $R_{\rm V}$ \mbox{($\equiv A_{\rm V}/E{\rm (B - V)}$)} (Cardelli et al. 1989). 
The parameter $R_{\rm V}$ can also serve as an indicator of galactic environmental characteristics, and its value in the dense region is larger than that in diffuse ISM (Whittet 2003).
In this work, the derived curve is not normalised, because the optical/UV SEDs of quasars are directly related to the fundamental plane of black hole activity, and the continuum slope is very different among different quasars.
We do not know the unreddened intrinsic continuum for a given quasar, and it is set to a quasar composite in the fitting. 
Obviously, there must be some deviation between them, and  it will be reflected in the parameter $c_{2}$. 
Thus, in the extinction measure process of this work, we cannot actually precisely extract the conventionally defined extinction parameters, such as $A_{\rm V}$, $E{\rm (B - V )}$ and $R_{\rm V}$, from the derived extinction curve.
Fortunately, the peak position and width of the bump feature are not correlated with one another (Fitzpatrick \& Massa 1986), but both are correlated with the slope of the extinction curve (Cardelli et al. 1989). Thus, the detection of the 2175 \AA~ bump feature from the derived extinction curve is viable, and we can investigate the extinction properties of these special dust grains.
The bump strength can be defined using these parameters: $A_{\rm bump} ~ = \pi c_{3}/2\gamma = 1.53\pm0.16$ $\mu$m$^{-1}$, which can be interpreted as rescaling the integrated apparent optical depth of the bump feature ($A_{\lambda}=\frac{2.5}{\rm ln 10} \tau_{\lambda}$).
Likewise, $A_{\rm bump}$ is also not normalised. In Gordon et al. (2003) and Fitzpatrick \& Massa (2007), the bump strength is defined as the normalised strength by the colour excess $(A_{\rm bump}/E{\rm (B-V)})$.

   \begin{figure}
   \centering
   \includegraphics[width=\hsize]{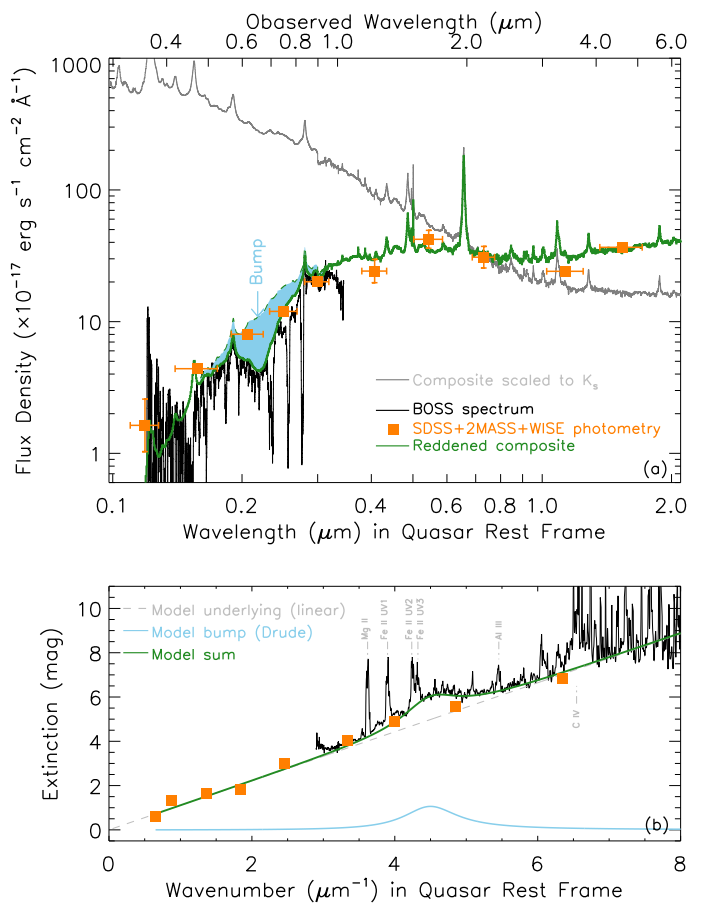}
   \caption{Panel (a): observed data of SDSS J1630+3119 and  2175 \AA~ bump feature in  spectrum.
The quasar spectra, including the original spectrum (black curve) and reddened quasar composites with the best-fitting extinction curves with/without the bump component (solid and dashed green curves), are overplotted for a comparison.
The sky blue region between them is the extinction bump of the MW-like dust.
The multiband magnitudes of SDSS J1630+3119 from the SDSS, 2MASS and WISE are presented by orange squares.
Panel (b): Observed extinction (orange squares and black curve) and  modelled extinction (green curve) of SDSS J1630+3119.
The grey dashed and sky blue solid lines present the underlying linear and bump components of the best-fitting extinction curve.
  The extinction curve rapidly decreases with an increasing wavelength into the infrared spectrum, and the extinction  is an unlimited approach to zero with $\lambda \to \infty$.} \label{fig3}
    \end{figure}

In Fig. 3(a), the best-fitting model for SDSS J1630+3119 is overplotted with the observed spectra  using a green solid cure.
To emphasise the requirement of the 2175 \AA~ bump feature in the extinction curve, we also overplot the reddened quasar composite with only the linear component of the best-fitting model (the green dashed curve). The difference between them corresponds to the bump extinction feature (sky blue shadow).
In Fig. 3(b), we present the observed (orange squares and black curve) and  modelled (green curve) extinction of SDSS J1630+3119.
The grey dashed and sky blue solid lines present the underlying and bump components of the best-fitting extinction curve.
One can find that the extinction curve rapidly decreases with an increasing wavelength into the infrared spectrum, and $E(x)$ is an unlimitd approach to zero with the a wavelength of $\lambda \to \infty$.

To confirm the existence of the 2175 \AA~  bump feature in SDSS J1630+3119, we employ several independent validation tests. The first is the simulation technique of the control sample developed by Jiang et al. (2010a,b).
This technique begins with the construction of a control sample, including 200 quasars with a minimal redshift differential relative to SDSS J1630+3119 and an $i'-$band spectral $S/N > 6$ from the SDSS DR7 quasar  catalogue.
Then, we fit each of them by reddening a quasar composite with a parameterised extinction curve. The parameters ($x_0$ and $\gamma$) are fixed to the best-fitting values of the 2175 \AA~ bump feature in SDSS J1630+3119.
Since the ``pseudo-continuum (the complex of nuclear continuum and Fe II emission multiplets)'' diversity between the quasar spectra and the composite can mimic  a false bump feature (e.g., Pitman et al. 2000), the bump strength distribution of the control sample can be considered as the measurement fluctuation of the 2175 \AA~  bump feature.  The distribution is expected to be a Gaussian profile, as shown by the pink curve in Fig. 4.
In principle, a bump candidate with a significance level of $> 3\sigma$ is considered real detection, where $\sigma$ is the standard deviation of the Gaussian profile
The bump strength in SDSS J1630+3119 is obviously  far from  the distribution ($\overline{A}_{\rm bump}=-0.01$ $\mu$m$^{-1}$ and $\sigma=0.05$ $\mu$m$^{-1}$), and the bump detection in SDSS J1630+3119 has   an extremely  high statistical significance.

\begin{figure}
\centering
\includegraphics[width=\hsize]{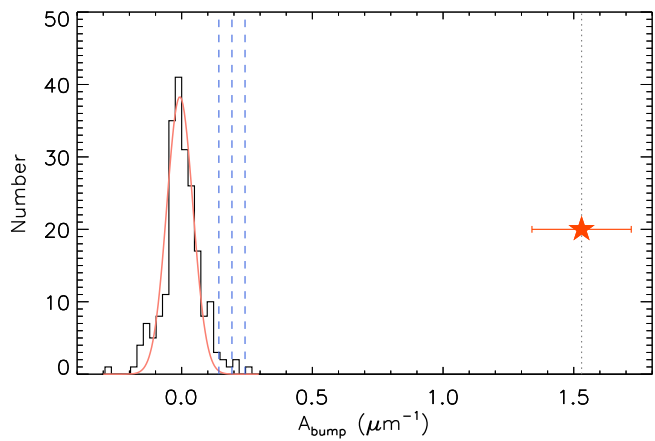}
\caption{Histogram of  fitted bump strengths of  control sample for SDSS J1630+3119. The pink solid line is the best-fitting Gaussian profile. Three vertical blue dashed lines indicate the 3$\sigma$, 4$\sigma$, and 5$\sigma$ boundaries of the Gaussian profile.  The red star indicates the bump strength derived in the spectrum of SDSS J1630+3119, which has an extremely high statistical significance level.} \label{fig4}
\end{figure}
The second test is more intuitionistic.   It focuses only on the bump features we are interested in.
For the purpose of comparison, we again use  the typical FeLoBALquasar, SDSS J1440+3710.
In Fig. 5, we present the corrected spectrum of SDSS J1630+3119 (sky blue curve) for bump reddening of the Drude component (sky blue solid curve in Fig. 3 (b)) and the observed spectra of SDSS J1630+3119 and SDSS J1440+3710.
The latter two are also shown in Fig. 2 for a comparison to identify BAL troughs.
In the figure, the corrected spectrum of SDSS J1630+3119 agrees well with the observed spectrum of SDSS J1440+3710 , except for the  two sides of the Fe II UV 1   troughs.
They share the same continuum, similar overlapping iron absorption lines and other metal absorption troughs.
In other words, the observed continuum difference between SDSS J1630+3119 and SDSS J1440+3710 is probably the 2175 \AA~ bump feature.

\begin{figure}
\centering
\includegraphics[width=\hsize]{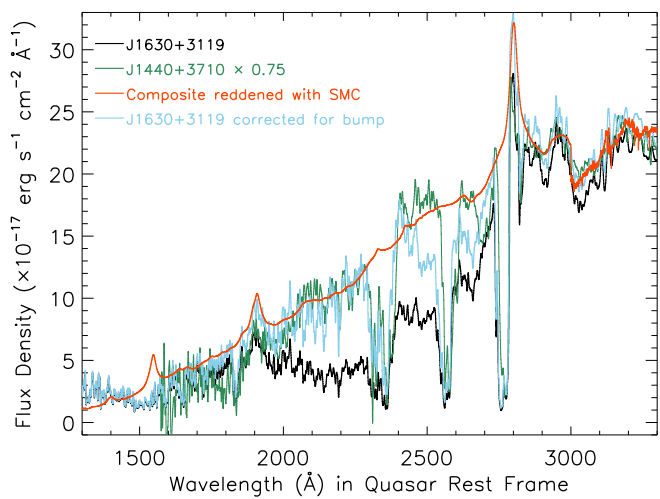}
\caption{Comparison of  corrected spectrum of SDSS J1630+3119  bump reddening (sky blue curve) and observed spectra of SDSS J1630+3119 and J1440+3710.
The corrected spectrum agrees very well with the observed spectrum of SDSS J1440+3710 (also shown in Fig. 2 by the green curve), and both spectra follow the reddened quasar composite (red curve) by the SMC extinction curve of $E{\rm (B-V)}=0.46$ very well.
The continuum difference between SDSS J1630+3119 and SDSS J1440+3710 is probably 2175 \AA~  bump feature.} \label{fig5}
\end{figure}

In the confirmation of 2175 \AA~  bump features  in BAL quasars (Zhang et al. 2015a), we performed continuum fitting by reddening a quasar composite with different extinction curves, namely, from the SMC, LMC2 Supershell (near the 30 Dor star-forming region; Gordon et al. 2003), LMC, and MW.
The bump is absent in the SMC extinction and gradually increases from the LMC2 Supershell to the LMC and the MW.
In Fig. 5 of this work, the reddened quasar composite by the SMC extinction curve of $E{\rm (B-V)}=0.46$ is present (red curve), and either the corrected spectrum of SDSS J1630+3119 or the observed spectrum of SDSS J1440+3710 could follow the composite continuum.
This implies that the extinction curves without the 2175 \AA~ bump feature are certainly not suitable for the original spectrum of SDSS J1630+3119.
For the extinction curves of the LMC2 Supershell, LMC, and MW, the addition of the 2175 \AA~ bump feature in the extinction curve would greatly improve the fitting results relative to the SMC curve. 

Considering that BAL quasars  have, on average, redder continua and stronger optical/UV Fe II emission multiplets than non-BAL quasars (e.g., Boroson \& Meyers 1992; Zhang et al. 2010), a few of the strongest Fe II and Fe III emission objects may be accidently identified as 2175 \AA~  bump features (Pitman et al. 2000). In Zhang et al. (2015a), we further mentioned trying to use the BAL composite to replace the quasar composite. We found that the fitting of the BAL composite generally shows slightly ($\sim 20 \%$) stronger bumps for all candidates.   
Meanwhile,  we also ruled out the possibility that the 2175 \AA~  bump feature is only an illusion of unusual Fe II and/or Fe III absorption troughs through the photoionization simulation of iron absorption. Although iron absorption multiplets are spread over the wavelength range of $2200 - 2800$ \AA, in either gas density ($n_{\rm H}$) -- column density ($N_{\rm H}$) -- ionization parameter ($U$) space, there are not enough iron transitions around 2175 \AA~ to produce wide and strong enough absorption troughs  that mimic the 2175 \AA~ bump feature.  More details can be found in Section 4 of Zhang et al. (2015a).

\section{Discussion}
  
SDSS J1630+3119 is not the only special case; in the appendices, we also present the analysis of 2175 \AA~ bump features observed toward two FeLoBAL quasars: SDSS J1020+6023 and SDSS J1349+3823.
The parameterised extinction curves yield prominent bump features at 2175 \AA~in the underlying extinction.
The bump parameters are $x_{0}=4.49\pm0.01$ $\mu$m$^{-1}$, $\gamma=1.20\pm0.03$ $\mu$m$^{-1}$, and $A\rm _{bump}=1.24\pm0.11$ $\mu$m$^{-1}$ for SDSS J1020+6023 and $x_{0}=4.67\pm0.01$ $\mu$m$^{-1}$, $\gamma=1.18\pm0.08$ $\mu$m$^{-1}$, and $A\rm _{bump}=2.68\pm0.25$ $\mu$m$^{-1}$ for SDSS J1349+3823.
Independent tests imply that both detections have extremely high statistical significance levels.
Compared with SDSS J1630+3119, the two cases appear to be the same, showing similar absorption troughs of Fe II multiplets, extreme reddening and prominent bump features; thus, they are listed as analogues of SDSS J1630+3119 in the appendices.
As Fig. A.1 and Fig. B.1 show, unlike SDSS J1630+3119, the spectra of SDSS J1020+6023 and SDSS J1349+3823 cover most of the bump profiles, 
except the blueward profile. 
The GALEX NUV magnitudes are high levels consistent with the reddened quasar composite, which suggests that the extinction shape and bump are well defined. 

\subsection{ Quasars associated or intrinsic systems?}

Usually, absorbers present in quasar spectra are classified as intervening or associated systems based on the blueshift velocity relative to the quasar rest frame. The former is produced by intergalactic clouds or galaxies located far from the background quasars, and the latter originates from materials associated with the host or intrinsically in quasar.  The 2175 \AA~ bump features in SDSS J1630+3119 and its analogues have the same redshifts as quasars, which indicates that these features are bound to be quasar-associated. 

  However, the carriers of these 2175 \AA~ bump features are still unable to be distinguished, whether the star formation region of the host or the nuclear region of the quasar exists. 
Although some works suggest that the profile and shift of the concomitant metal absorption lines might be able to be a discriminating pointer (e.g., Shi et al. 2020), that undoubtedly gives a higher   requirement for spectroscopic observations. In this work, the spectral resolution is far from achieving the degree of resolving   line profiles in absorption troughs; meanwhile, there is no narrow emission line in the wavelength coverage to measure the precise system redshift. We might be able to try to explore the origin of bump features by means of the simultaneous occurrence of BALs.
On the one hand, the host of quasars may suffer strong high-energy radiation from the nucleus,  while the quasar feedback on hosts may also produce a disturbance in the ISM. Hence, the probability of 2175 \AA~ bump features existing in hosts is relatively  small compared to quiescent galaxies, which is why quasar-associated 2175 \AA~  bump features  are rarely found.
On the other hand, previous statistical works indicate that  approximately  15\% of optically selected quasars show BAL features, and the detection probabilities of LoBALs and FeLoBALs are two and three powers of the above number (e.g., Weymann et al. 1991; Reichard et al. 2003; Trump et al. 2006; Gibson et al. 2009; Zhang et al. 2010). However, the cases found in  the literature  (Zhang et al. 2015a; Pan et al. 2017; Shi et al. 2020) and this work give a different conclusion. From these works, we collected 28 sources whose 2175 \AA~  bump features   have the same redshifts as quasars. Thus, the BAL fractions are nearly 36\% (BALs), 29\% (LoBALs), and 11\% (FeLoBALs) in this small sample. This situation suggests that there is a strong correlation between  occurrences  of BALs and 2175 \AA~ bump features, and that the 2175 \AA~  bump features  we observed are intrinsic to quasars rather than the hosts.

Admittedly, this small sample has a strong selection effect. The exploration in Zhang et al. (2005a) is based on the SDSS DR10 Mg II absorption-line sample, in which the quasars with unusual absorption lines (i.e., FeLoBALs) are removed through visual inspection. However,  this work begins  from the known SDSS DR7 FeLoBAL quasars; therefore,  the fraction of the FeLoBAL-2175 \AA~  bump features  is relatively grossly underestimated.
Moreover, statistical studies of optically selected BAL quasars indicate that LoBAL quasars show a moderate level of reddening and FeLoBAL quasars are heavily reddened (e.g., Hall et al. 2002; Reichard et al. 2003; Zhang et al. 2010; Dunn et al. 2015), and the additional 2175 \AA~ bump features make these objects fainter and easier to miss in the magnitude-limited sample. If an unbiased systematic exploration for 2175 \AA~ bump features is performed in homogeneous quasar samples, it is believed that the foregoing conclusion will be greatly enhanced.

\subsection{Comparison of the bump parameters}

\begin{figure*}
\centering
\includegraphics[width=\hsize]{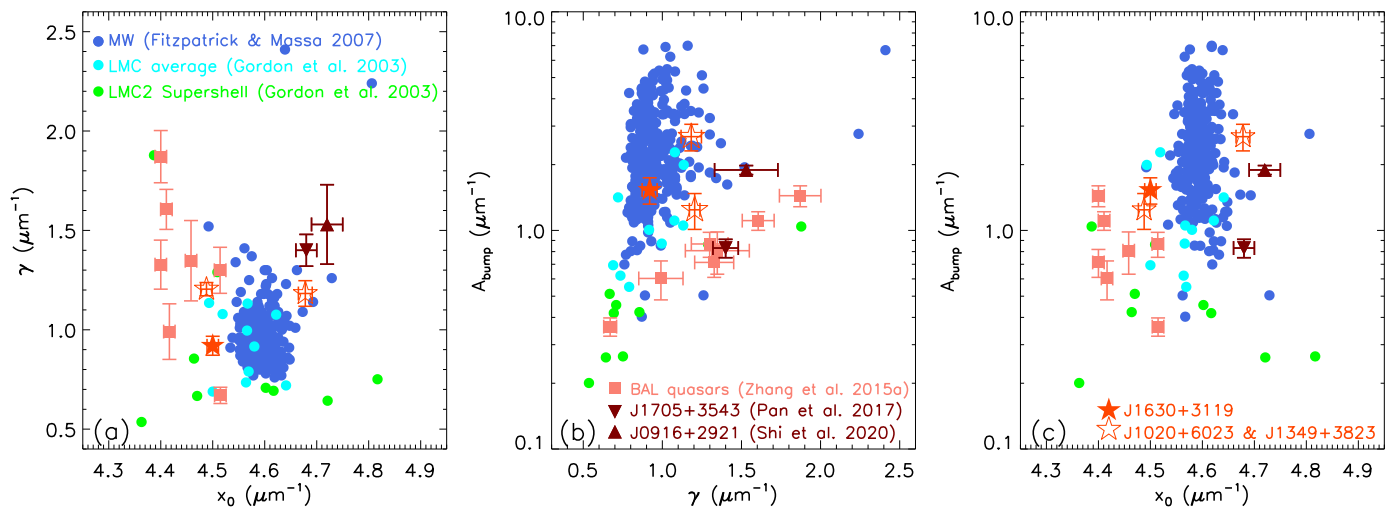}
\caption{  Comparison of  bump parameters for  MW (blue circles), LMC average (navy circles), and LMC2 Supershell (green circles); BAL quasars (pink squares) and two cases in  literature (brown triangles) with intrinsic 2175 \AA~ bump features, and SDSS J1630+3119 (red filled stars). For visual clarity, the error bars on the MW, LMC average and LMC2 Supershell bumps are not displayed. The bump in SDSS J1630+3119 is relatively closer to those in MW, and they share similar absorption strength and width, but the bump peak in SDSS J1630+3119 is located at a longer wavelength than the galactic interstellar curves. By contrast, the bumps in the LMC average, LMC2 Supershell, BAL quasars and the other two cases present broader absorption features with more disordered peak positions and weaker absorption strengths. The results of the two analogues are also shown by red open stars.} \label{fig6}
\end{figure*}

  Fig. 6 illustrates the bump parameters of the 2175 \AA~ bump features in SDSS J1630+3119 and its analogues. For a  comparison, the panels also include those in the Local Group and  the other 9 quasar intrinsic 2175 \AA~  bump features. Among the 12 quasar intrinsic cases, 3 objects (SDSS J0916+2921, SDSS J1020+6023 and SDSS J1349+3823) have redshifts of $z \sim 1.0$,  and the others have higher redshifts ($z\sim 2.0$). The absorbers in the MW (blue circles) have a large absorption strength ($A_{\rm bump}=2.48\pm1.15$ $\mu$m$^{-1}$) and a constant peak position ($x_{0}=4.59\pm0.027$ $\mu$m$^{-1}$), and their widths vary  over  a wide range of $\gamma=1.0\pm0.25$ $\mu$m$^{-1}$.
The 2175 \AA~  bump features  in BAL quasars (pink squares; Zhang et al. 2015a) present broader bump features at smaller wavenumbers ($\gamma = -1.3\pm 0.39$ $\mu$m$^{-1}$, $x_0 = 4.45\pm 0.05$ $\mu$m$^{-1}$), and their strengths are much weaker than those of the galactic bumps at 2175 \AA~ on average, but similar to those of the LMC average (navy circles) and LMC2 Supershell (green circles) bumps.
However, two 2175 \AA~ bump features (brown triangles) reported in Pan et al. (2017) and Shi et al. (2020) are relatively prone to the MW sightlines that deviate most from the main distribution. The 2175 \AA~ bump features in SDSS J1630+3119 and its analogues are significantly stronger than the previous detections in BAL quasars and SDSS J1705+ 3543. Their strengths are very close to the averaged strength of the galactic bumps.
For the bump width, the  bump in SDSS J1630+3119 has almost the same average width as those in the MW, and the    bumps in the two analogues  are slightly wider, but they are not out of range of those in the MW. For the peak position, only in one case (SDSS J1349+3823),  the bump peaks at $2141\pm4.5$ \AA, the blueward of 2175 \AA ; the bumps  in another two cases (SDSS J1020+6023 and SDSS J1630+3119) tend to peak at  larger wavelengths, and their peak positions are $2227\pm4.9$ \AA~ and $2217 \pm 5.0$ \AA, respectively.

The  bump peak position seems to vary in different studies. We note that this parameter depends on the accuracy of the redshift measurement, which is a weakness of BAL quasars. In their spectra, there is a lack of narrow emission lines and not enough complete BEL profiles to measure the precise redshift.  The position deviation of the bumps in all BAL quasars is $0.08$ $\mu$m$^{-1}$, and this value corresponding to a redshift is approximately 0.018. Although there are many disadvantages in the redshift measurement of BAL quasars, the SDSS redshift accuracy is much higher than the above value. This suggests that the redshift uncertainty is not sufficient to cause such a large variation. Moreover, none of the 2175 \AA~  bump features  in this work and other quasar intrinsic cases are out of range of the LMC average and LMC2 Supershell bumps  (Gordon et al. 2003). Thus, we tend to think that the different parameter distributions of the 2175 \AA~  bump features are intrinsic and  are possibly attributed to variations in the grain-size and/or chemical composition of the dust (Draine 2003).

\subsection{Survival  environment of 2175 \AA~dust grains}

  For these FeLoBAL quasars with 2175 \AA~ bump features, which structure do 2175\AA~ dust grains survive in: the BAL region, or something else?   In the common sense, outflows may emerge from the outer region of the accretion disk or even the innermost region of the torus, in which the gas clouds are dusty and relatively cold. These dusty clouds are uplifted and exposed to the central engine.
Generally, the low density part is rapidly heated and highly ionised, and the dust in this would be sublimated.  The dust grains survive in dense regions and may even form in dense clouds embedded in outflows (Elvis et al. 2002).
However, this does not mean that 2175 \AA~ dust grains must exist in the outflows. Nearly two-thirds of quasars with quasar-associated 2175 \AA~ bump features are non-BAL quasars. When the sightline to continuum source intersects the outflow clouds,  the 2175 \AA~  bump feature and BAL troughs would coexist in the spectrum, which originate from the dust grains and ionised gases carried by outflows, respectively.
  In addition,  the torus may be the primary dust warehouse in the nuclear region of quasars. Large amounts of dust grains gather here, and the stacked structure provides a good shelter for them. The absorption features of the silicate grains at 9.7 $\mu$m and the polycyclic aromatic hydrocarbon (PAH) molecules at 3.4 $\mu$m are collected in the mid-IR spectra of many Type II Seyferts.  
We suspect that such dust is widespread; it also certainly requires a proper orientation angle for 2175 \AA~bump features in quasars; otherwise, the dust grains either do not obscure the central engine or are completely obscure, and then the source is transformed into Type II.

In Zhang et al. (2015a), we proposed another probable picture taken from the early evolution of quasars. The AGN activity destroys and clears the dust grains in the surrounding space of  the nucleus, which are prepared through the persistent nucleosynthesis processes of carbon, and the formation of dust occurs throughout the evolution of stars and galaxies before the black hole is ignited. Outflow winds shield high-energy photons from the central engine and extend the survival period of dust in the nuclear region of quasars.

  How these dust grains enter into (or are assembled) and survive in the nuclear region of quasars and the survival environment and the dust grain itself are interesting questions  that deserve further observational study and investigation.
For example, follow-up high-resolution spectroscopy at large telescopes will present abundant gas absorption lines in association with dust, which could be used to measure the chemical abundance and dust depletion, obtain the column density of hydrogen and the dust-to-gas ratio, and even detect neutral and molecular   gases, finally understanding  the chemical and physical conditions in absorber systems.

Furthermore, although the origin of the 2175 \AA~  bump features  remains unclear, carbonaceous dust has long been proposed as a candidate. Laboratory and theoretical studies confirmed that materials such as graphite, aromatic hydrocarbon molecules, fullerene, hydrogenated amorphous carbon, and even carbon nanotubes  could produce a broad extinction bump at approximately 2175 \AA~ (e.g., Stecher \& Donn 1965; Wright 1988; Mennella et al. 1998; Aannestad 1992; Joblin et al. 1992).  The carrier of the 2175 \AA~ bump features is probably one of them or a mixture of several materials. Follow-up near-/mid-IR spectroscopy will be expected to present the corresponding absorption and emission features of the materials.

\section{Summary}
In this paper,  we reported for the first time the discovery of 2175 \AA~ bump features  in the optical spectra of the FeLoBAL quasars SDSS J1630+3119 and its two  analogues. They are typical FeLoBAL quasars, and their optical spectra present numerous absorption troughs from high- and low-ionization species, such as C IV, Al III, Mg II and Fe II multiplets. 
After masking these broad absorption features, parameterised extinction curves were used to redden the quasar composite to fit the observed spectra. The UV to near-IR broadband SEDs of these objects verify the efficiency of spectral continuum fitting.
The fitting yields prominent extinction bumps at 2175 \AA, which are also confirmed again through several independent validation tests.
Compared to the bump features in the LMC and LMC2 Supershell, the detections in this work are much closer to those in the MW.
Our 2175 \AA~ bump features are detected in the quasar rest frame.  Taking the fractions of high- and (iron) low-ionization BALs into account,  we consider that they, as well as those in the BAL quasars of Zhang et al. (2015), are intrinsic to quasars rather than the quasar host and intervening galaxies.
These 2175 \AA~  bump features may be related to the bumps seen in the Local Group and may be a counterpart of the 2175 \AA~  bump features  under different conditions in the quasar environment.
  Traversing the various structures of AGNs, we suggest that only a dusty torus can provide a suitable context for the survival of 2175 \AA~ bump carriers. 
The dust grains perhaps survive  in the dusty torus, and the shielding effect of outflow clouds allows the MW-like dust to be assembled or to extend the survival period in the nuclear region. 
  To  further investigate the intrinsic nature of FeLoBAL-2175 \AA~bump feature, we need follow-up monitoring of absorption lines and mid-infrared spectroscopy to study PAH emissions.

\begin{acknowledgements}
We thank the anonymous referee for their  constructive comments and suggestions.
This work is supported by the National Natural Science Foundation of China under  Grant No. NSFC-12173026  and the  Natural Science Foundation of Shanghai under Grant No. 20ZR1473600.
This publication makes use of data products from the Two Micron All Sky Survey, which is a joint project  of the University of Massachusetts and the Infrared Processing and Analysis Center/California Institute of Technology, funded by the National Aeronautics and Space Administration and the National Science Foundation.
This publication makes use of data products from the Wide-field Infrared Survey Explorer, which is a joint project of the University of California, Los Angeles, and the Jet Propulsion Laboratory/California Institute of Technology, funded by the National Aeronautics and Space Administration.
This research is based on observations made with the Galaxy Evolution Explorer, obtained from the MAST data archive at the Space Telescope Science Institute, which is operated by the Association of Universities for Research in Astronomy, Inc., under NASA contract NAS 5-26555.
The CSS survey is funded by the National Aeronautics and Space Administration under Grant No. NNG05GF22G issued through the Science Mission Directorate Near-Earth Objects Observations Program.  The CRTS survey is supported by the U.S.~National Science Foundation under grants AST-0909182.
The Pan-STARRS1 Surveys (PS1) and the PS1 public science archive have been made possible through contributions by the Institute for Astronomy, the University of Hawaii, the Pan-STARRS Project Office, the Max-Planck Society and its participating institutes, the Max Planck Institute for Astronomy, Heidelberg and the Max Planck Institute for Extraterrestrial Physics, Garching, The Johns Hopkins University, Durham University, the University of Edinburgh, the Queen's University Belfast, the Harvard-Smithsonian Center for Astrophysics, the Las Cumbres Observatory Global Telescope Network Incorporated, the National Central University of Taiwan, the Space Telescope Science Institute, the National Aeronautics and Space Administration under Grant No. NNX08AR22G issued through the Planetary Science Division of the NASA Science Mission Directorate, the National Science Foundation Grant No. AST-1238877, the University of Maryland, Eotvos Lorand University (ELTE), the Los Alamos National Laboratory, and the Gordon and Betty Moore Foundation.
ZTF is supported by the National Science Foundation and a collaboration including Caltech, IPAC, the Weizmann Institute for Science, the Oskar Klein Center at Stockholm University, the University of Maryland, Deutsches Elektronen-Synchrotron and Humboldt University, Lawrence Livermore National Laboratory, the TANGO Consortium of Taiwan, the University of Wisconsin at Milwaukee, Trinity College Dublin, and Institut national de physique nucleaire et de physique des particules. Operations are conducted by COO, IPAC and University of Washington.
The part of data presented herein were obtained at the W. M. Keck Observatory, which is operated as a scientific partnership among the California Institute of Technology, the University of California and the National Aeronautics and Space Administration. The Observatory was made possible by the generous financial support of the W. M. Keck Foundation.  The authors wish to recognize and acknowledge the very significant cultural role and reverence  that the summit of Mauna Kea has always had within the indigenous Hawaiian community. We are most fortunate to have the opportunity to conduct observations from this mountain.
Funding for SDSS-III has been provided by the Alfred P. Sloan Foundation, the Participating Institutions, the National Science Foundation, and the U.S. Department of Energy Office of Science. The SDSS-III web site is http://www.sdss3.org/.
SDSS-III is managed by the Astrophysical Research Consortium for the Participating Institutions of the SDSS-III Collaboration including the University of Arizona, the Brazilian Participation Group, Brookhaven National Laboratory, Carnegie Mellon University, University of Florida, the French Participation Group, the German Participation Group, Harvard University, the Instituto de Astrofisica de Canarias, the Michigan State/Notre Dame/JINA Participation Group, Johns Hopkins University, Lawrence Berkeley National Laboratory, Max Planck Institute for Astrophysics, Max Planck Institute for Extraterrestrial Physics, New Mexico State University, New York University, Ohio State University, Pennsylvania State University, University of Portsmouth, Princeton University, the Spanish Participation Group, University of Tokyo, University of Utah, Vanderbilt University, University of Virginia, University of Washington, and Yale University.

\end{acknowledgements}

\begin{appendix}

\section{SDSS J1020+6023}

SDSS J1020+6023 was selected as a candidate of  the intrinsic 2175\AA~bump feature according to the SDSS spectrum. Its broad extinction bump was confirmed by  UV-to-IR broadband SED and again by a  follow-up Keck spectroscopy.

SDSS J1020+6023 was imaged at five photometric bands on March 3, 2000, and targeted as a quasar candidate based on its location in the $u'g'r'i'$ colour cube (Richards et al. 2004) for spectroscopic observations by the SDSS survey. Two observations were taken  on  March 05, 2002, and the spectrum we used was extracted from the SDSS DR7. 
This object did not vary significantly between the SDSS photometric and spectroscopic observations, since its spectrum is in excellent agreement with photometric data.
SDSS J1020+6023 is also a typical FeLoBAL quasar, classified into several existing BAL catalogs (Trump et al. 2006; Gibson et al. 2009; Zhang et al. 2010; Shen et al. 2011).
The blue band spectrum shows overlapping iron absorption lines (2300-2415\AA, 2550-2630\AA~and 2700-2770\AA) and Mg II broad absorption  troughs.
Except for  iron emission/absorption multiplets and narrow absorption lines of Mg I and Ca II H/K, no other emission or absorption lines can be identified from the spectrum.

A follow-up spectroscopic observation was obtained  on March 8, 2013,  on the 10 m Keck II telescope using the Echellette Spectrographs and Imager (ESI; Sheinis et al. 2002).
The high resolution ($R\sim30,000$), high signal-to-noise ratio (median~$S/N\approx11$) spectrum  was  determined to distinguish the structure of absorption features.
We implemented the $0.75\arcsec$ slit to match the seeing, where one exposure of 1200 seconds was taken. The data were reduced and calibrated using the ESIRedux software package.
The entire spectra cover the full optical range from 3900 \AA~to 1.1 $\mu$m. It is  highly  consistent with the SDSS spectrum and $r'$-, $i'$- and $z'$-band photometric data.
  This high-resolution spectroscopic observation is  widely used in extinction curve fitting and all measurements of broad and narrow absorption lines.

SDSS J1020+6023 is very red ($\Delta (g-i)=1.5180$) with a dramatic flux drop shortward of Mg II BAL, suggesting the presence of heavy dust extinction.
We also collected broadband photometric data of SDSS J1020+6023 from other available large surveys at UV wavelengths (the Galaxy Evolution Explorer or GALEX; Martin et al. 2005) and IR wavelengths (the 2MASS and WISE).
The IR and optical photometric data join smoothly, suggesting that the quasar did not vary significantly between the 2MASS, WISE and SDSS observations. The broadband SED broken  at the SDSS $r$-band again  confirmed heavy dust extinction.
After correcting for galactic reddening using the dust map of Schlegel et al. (1998) and the Fitzpatrick (1999) reddening curve, we transformed the magnitudes and optical spectrum into a rest frame with a quasar redshift.
The same fitting process was used for exploring the 2175 \AA~ bump feature in SDSS J1020+6023.
The best-fitting model yields $c_{1} = -1.82\pm0.02$, $c_{2}=0.54\pm0.01$, $c_{3}=0.95\pm0.08$, $x_{0}=4.49\pm0.01$ $\mu$m$^{-1}$, and $\gamma=1.20\pm0.03$ $\mu$m$^{-1}$.
The strength of the  2175 \AA~ bump feature is $A\rm _{bump}=1.24\pm0.11$ $\mu$m$^{-1}$.

  This  model is consistent with multiband photometry with effective wavelengths less than the $H$-band.
The GALEX NUV magnitude is high, consistent with the reddened quasar composite spectrum.  The  $u'$-band is affected by the expected strong \mbox{Al III} absorption and has  a lower radiation flux than the model spectrum.
  However,  the near-IR photometric fluxes ($H$, $K_{s}$, and $W1$) of SDSS J1020+6023 are under the reddened spectral template.
The reason is that the near-IR composite was constructed from observations of 27 luminous, low-redshift, UV-excess selected quasars, and it was excellently described by the combined model with a single power-law slope and a black body from H$\alpha$ to 3.5 $\mu$m (Glikman et al. 2006).
This result strongly suggests the presence of significant quantities of hot dust in the near-IR composite.
We added a dust emission component with a temperature of $\sim 1700$ K to near-IR magnitudes of SDSS J1020+6023, and the correctional fluxes (open squares) are  extremely consistent with the near-IR composite.
The typical dust sublimation temperature is $\sim 1500 - 2000$ K (e.g., Tuthill et al. 2001; Monnier \& Millan-Gabet 2002), and the extra hot dust in this object is probably the inner edge of the torus.
The combination spectrum together with the IR through optical to UV photometric data, the best-fitting model and the  observed/modelled  extinction curves are  presented  in Fig. A.1.

Independent tests were  employed  to check the authenticity of the detection. As shown in Fig. A.2, the bump strength distribution of the control sample peaks at $\overline{A}_{\rm bump}=0.13$ $\mu$m$^{-1}$ with a variance of $\sigma=0.16$ $\mu$m$^{-1}$, and the bump feature in SDSS J1020+6023 is detected at $> 5\sigma$.
In Fig. A.2 (b), the scaled observed spectrum of the FeLoBAL quasar SDSS J112828.31+011337.9  is overplotted for a  comparison.
This case has almost the same Mg II and Fe II absorption troughs as SDSS J1020+6023, and its spectrum agrees well with the corrected spectrum of SDSS J1020+6023.
Both spectra follow the reddened quasar composite by the SMC extinction curve of $E{\rm (B-V)}=0.25$.

\begin{figure}
\centering
\includegraphics[width=\hsize]{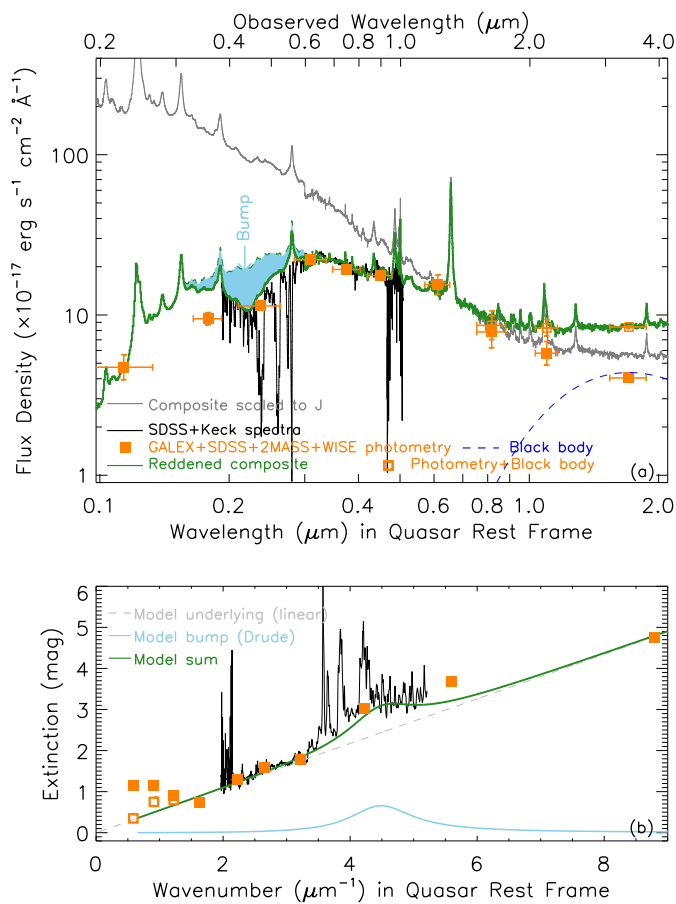}
\caption{Same as Fig. 3 but for SDSS J1020+6023.
A dust emission with $T \sim 1700$ K (blue dashed line) is added to the near-IR magnitudes  to better match  the composite, since the NIR composite from Glikman et al. (2006) has a stronger IR emission than SDSS J1020+6023. The correctional multiband magnitudes of SDSS J1020+6023 from the 2MASS and WISE surveys are presented by orange open squares. } \label{figa1}
\end{figure}

\begin{figure}
\centering
\includegraphics[width=\hsize]{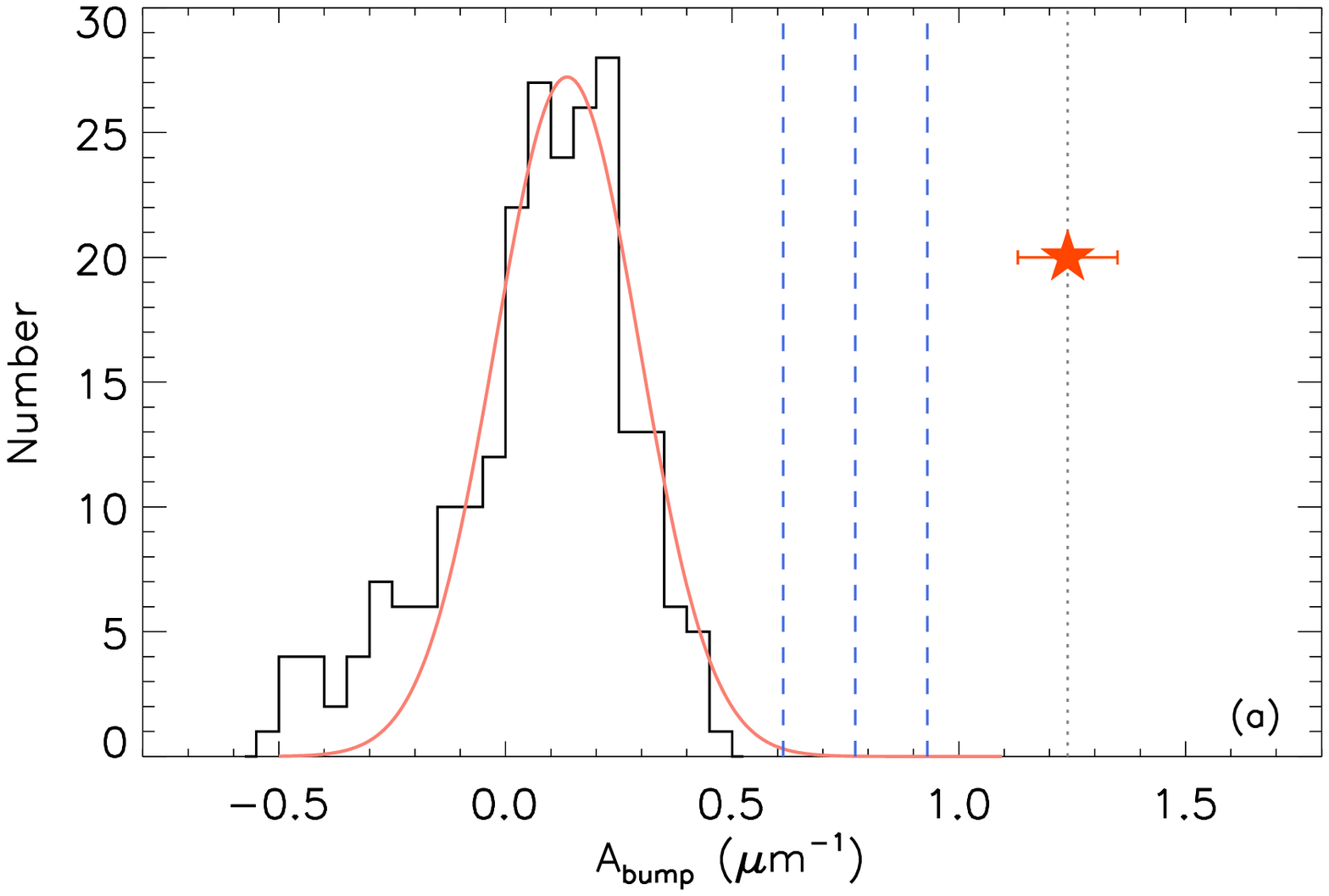}
\includegraphics[width=\hsize]{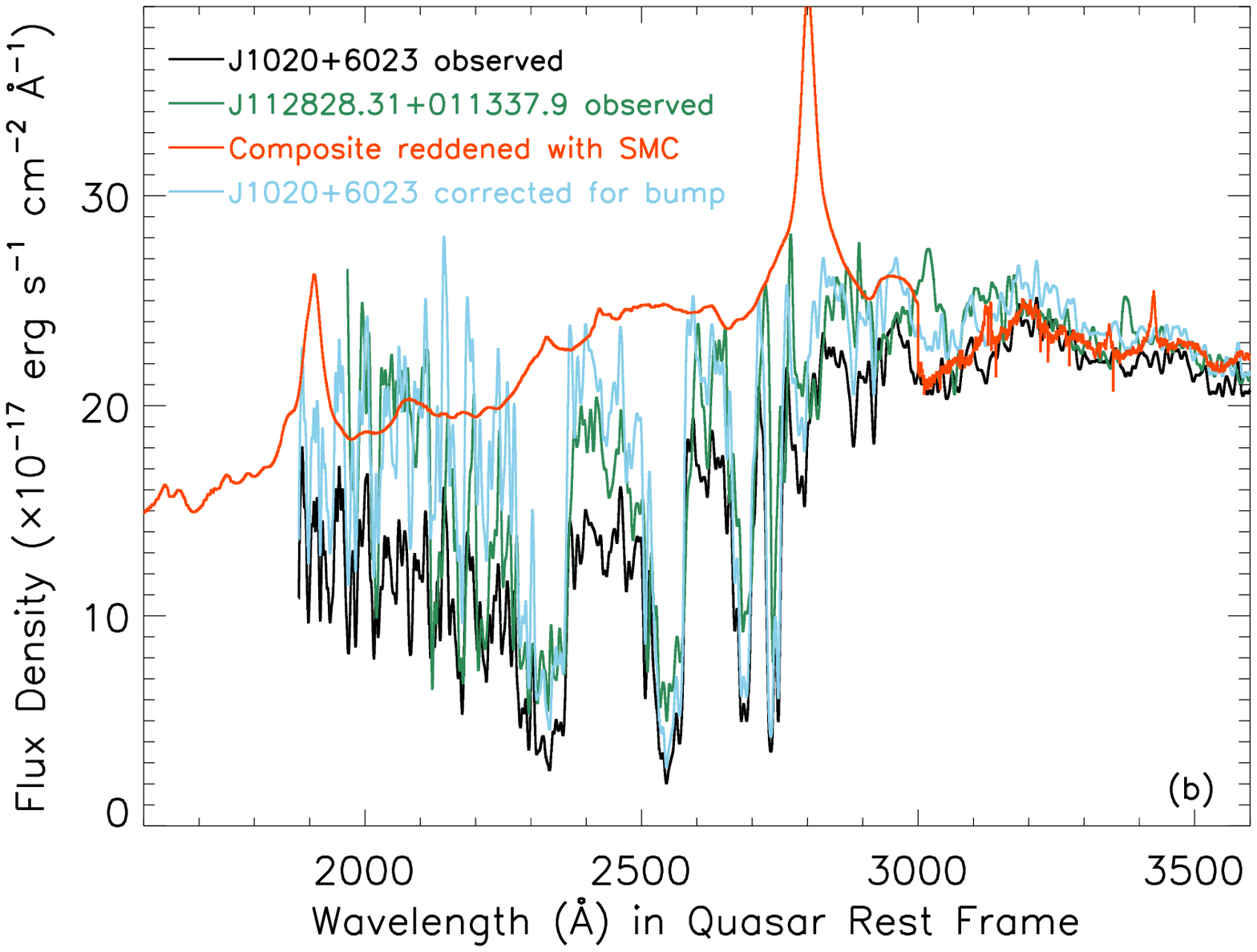}
\caption{Same as Fig. 4 and Fig. 5, but for SDSS SDSS J1020+6023. The  continuum  slopes of the corrected spectrum of SDSS J1020+6023 and the observed spectrum of SDSS J112828.31+011337.9 follow the reddened quasar composite (red curve) by the SMC extinction curve of $E{\rm (B-V)}=0.25$.} \label{figa2}
\end{figure}

\section{SDSS J1349+3823}

SDSS J1349+3823 is another  analogue of SDSS J1630+3119. The archives provide multiwavelength photometric magnitudes from the GALEX, SDSS and 2MASS surveys and spectroscopic fluxes from the SDSS survey. Its 1D spectrum from 3800 \AA~ to 9200 \AA~ was accessed from the SDSS DR7. After correcting for galactic reddening using the dust map of Schlegel et al. (1998) and the Fitzpatrick (1999) reddening curve, we transformed the magnitudes and optical spectrum into a quasar rest frame. The photometric and spectroscopic data are shown  using  orange squares and black curves in Fig. B.1 (a).
The optical spectrum simultaneously shows  a  remarkable intrinsic 2175 \AA~  bump feature  and broad absorption troughs of Mg II and Fe II multiplets UV 1, UV 2, UV 3 and UV 62,63. The extremely low flux at the $u'$ band implies that there are possible strong absorption troughs of Al III and even Fe III UV 34 $\lambda\lambda\lambda$1895.46, 1914.06, and 1926.30, and UV 48 $\lambda\lambda\lambda$2062.21, 2068.90, and 2079.62.

The same fitting process was used  to explore the 2175 \AA~ bump feature in SDSS J1020+6023.
The best-fitting model yields $c_{1}=-0.73\pm0.02$, $c_{2}=0.37\pm0.01$, $c_{3}=2.02\pm0.13$, $x_{0}=4.67\pm0.01$ $\mu$m$^{-1}$, and $\gamma=1.18\pm0.08$ $\mu$m$^{-1}$  , where the bump strength is $A\rm _{bump}=2.68\pm0.25$ $\mu$m$^{-1}$.
This case,  similar to SDSS J1020+6023, has a low redshift of $\sim 1.0$; therefore,  the SDSS spectrum cannot cover the shortward of the 2175 \AA~ bump feature.
However, the GALEX NUV and 2MASS near-IR photometric data  fall right on the modelled spectrum, which suggests that the shape and bump of the extinction curve are well defined.
In Fig. B.1 (a) and (b), the best-fitting model and  observed/modelled extinction curves are  presented. We also performed  tests as for J1630+3119 to confirm the bump detection in SDSS J1349+3823. As shown in Fig. B.2, the bump feature in SDSS J1349+3823 is detected at $\gg 5\sigma$. The corrected spectrum of SDSS J1349+3823 agrees with the observed spectrum of the FeLoBAL  quasar SDSS J084044.41+363327.8, and their spectral continua follow the reddened quasar composite by the SMC extinction curve of $E{\rm (B-V)}=0.17$.
\begin{figure}
\centering
\includegraphics[width=\hsize]{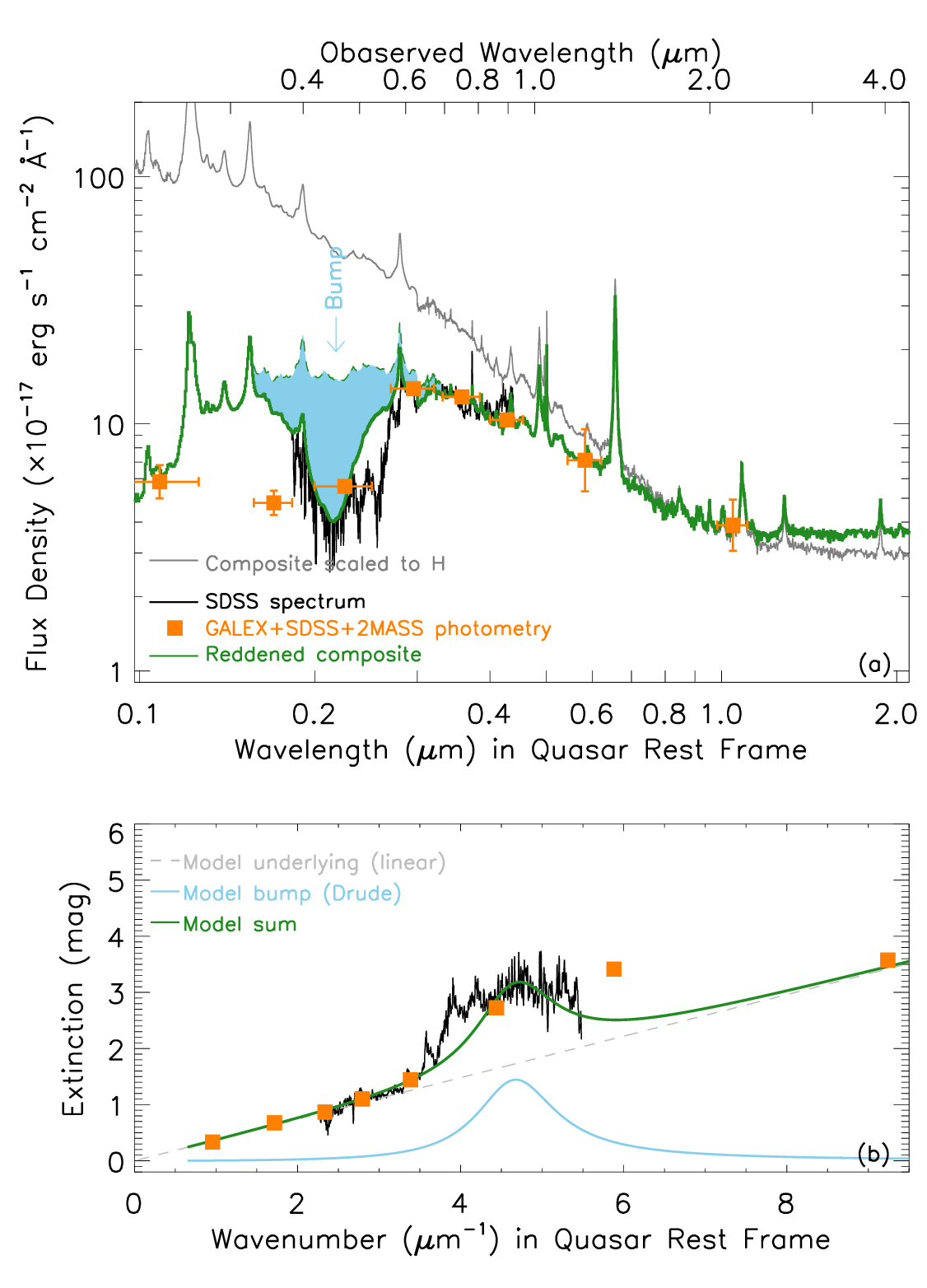}
\caption{Same as Fig. 3 but for SDSS J1349+3823. } \label{figb1}
\end{figure}

\begin{figure}
\centering
\includegraphics[width=\hsize]{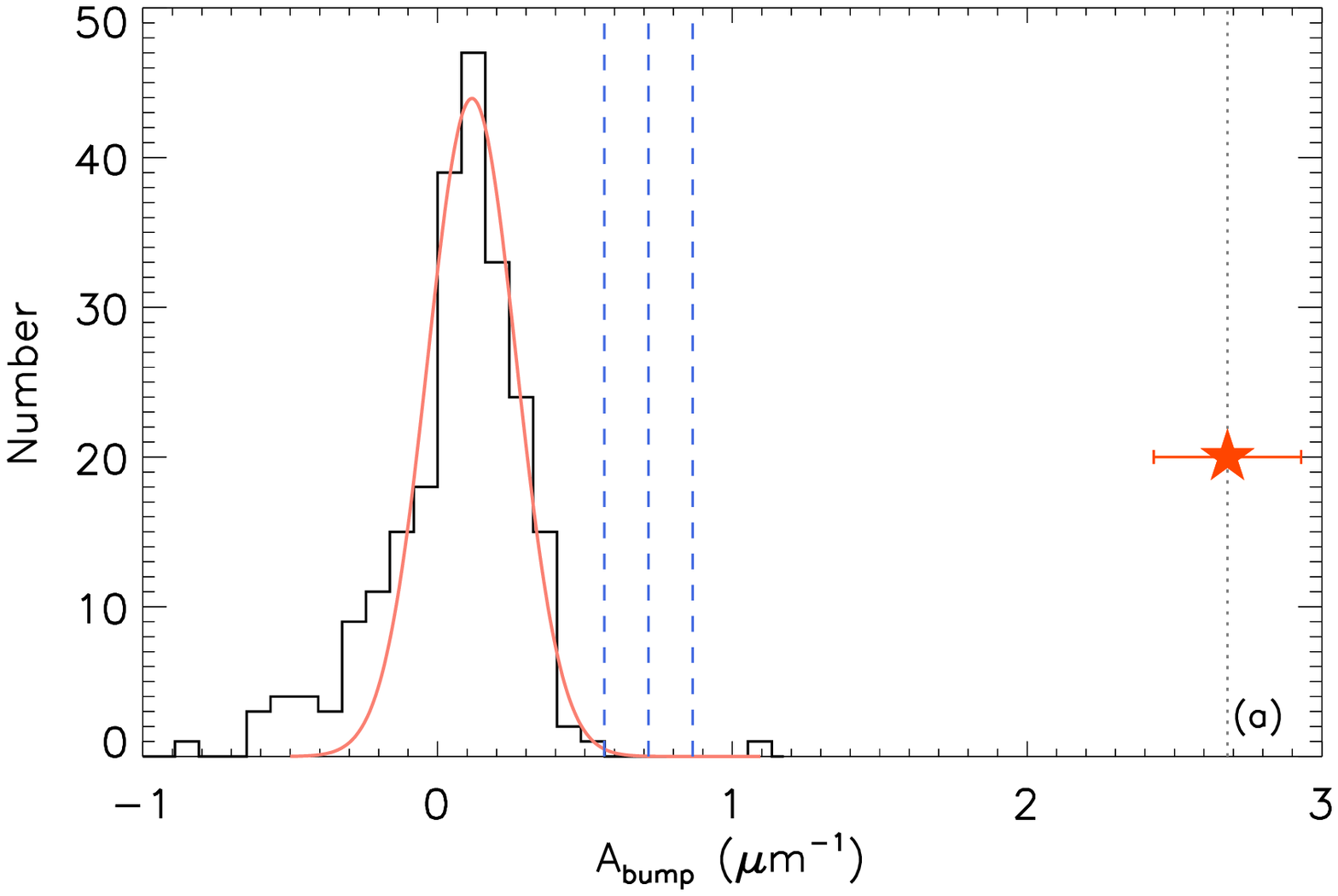}
\includegraphics[width=\hsize]{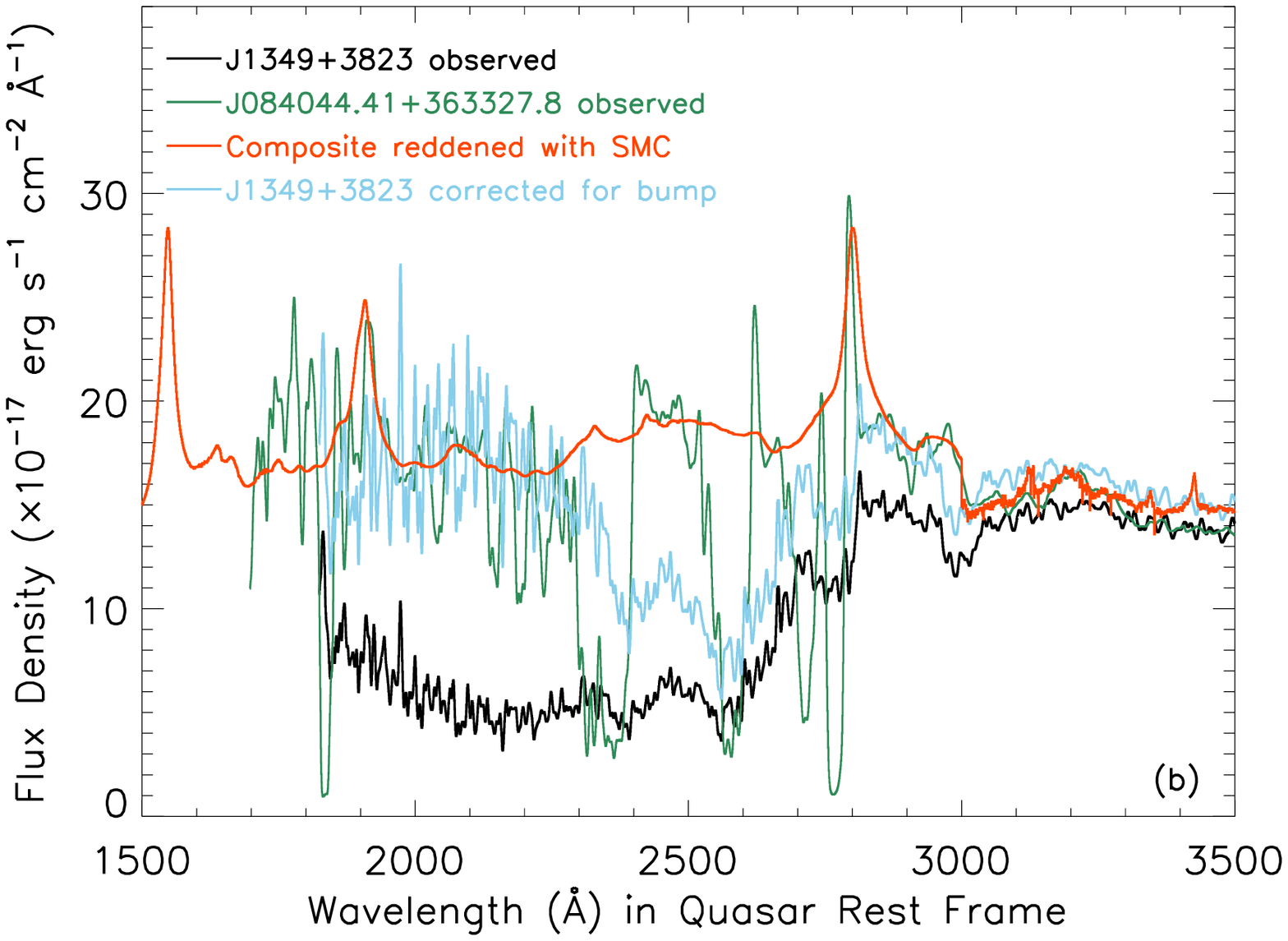}
\caption{Same as Fig. 4 and Fig. 5, but for SDSS J1349+3823. The  continuum slopes of the corrected spectrum of SDSS J1349+3823 and observed spectrum of SDSS J084044.41+363327.8 follow the reddened quasar composite (red curve) by the SMC extinction curve of $E{\rm (B-V)}=0.17$.} \label{figb2}
\end{figure}

\end{appendix}

\end{document}